\documentclass[sigconf, nonacm]{acmart}

\newcommand\vldbavailabilityurl{https://github.com/lakreis/GPU-RMQ}

\newcommand\RTXRMQauthors{Meneses et al. }

\usepackage{amsmath}
\usepackage{graphicx}
\usepackage{xcolor}
\usepackage{xfrac}
\usepackage{listings}%
\usepackage{subcaption}
\usepackage[english]{babel}
\usepackage{hyperref}
\addto\extrasenglish{%
}

\newcommand{\ours}{\textbf{GPU-RMQ}}
\newcommand{\oursCoalesced}{\textbf{GPU-RMQ (CL)}}
\newcommand{\oursVector}{\textbf{GPU-RMQ (VL)}}
\newcommand{\exhaustive}{\textbf{Full GPU Scan}}
\newcommand{\LCA}{\textbf{LCA}}
\newcommand{\RTXRMQ}{\textbf{RTXRMQ}}
\newcommand{\HRMQ}{\textbf{HRMQ}}

\begin{document}
\title{GPU-RMQ: Accelerating Range Minimum Queries on Modern GPUs}


\author{Lara Kreis}
\affiliation{%
  \institution{Johannes Gutenberg University}
  \city{Mainz}
  \state{Germany}
  \postcode{43017-6221}
}
\email{lakreis@students.uni-mainz.de}

\author{Justus Henneberg}
\orcid{0000-0002-1825-0097}
\affiliation{%
  \institution{Johannes Gutenberg University}
  \city{Mainz}
  \country{Germany}
}
\email{henneberg@uni-mainz.de}

\author{Valentin Henkys}
\orcid{0009-0002-9379-7129}
\affiliation{%
  \institution{Aperio Space Technologies}
  \city{Dieburg}
  \country{Germany}
}
\email{valentin.henkys@aperiospace.com}

\author{Felix Schuhknecht}
\affiliation{%
  \institution{Johannes Gutenberg University}
  \city{Mainz}
  \country{Germany}
}
\email{schuhknecht@uni-mainz.de}

\author{Bertil Schmidt}
\affiliation{%
  \institution{Johannes Gutenberg University}
  \city{Mainz}
  \country{Germany}
}
\email{bertil.schmidt@uni-mainz.de}

\begin{abstract}
Range minimum queries are frequently used in string processing and database applications including biological sequence analysis, document retrieval, and web search. 
Hence, various data structures have been proposed for improving their efficiency on both CPUs and GPUs.
Recent work has also shown that hardware-accelerated ray tracing on modern
NVIDIA RTX graphic cards can be exploited to answer range minimum queries by expressing queries as rays, which are fired into a scene of triangles representing minima of ranges at different granularities. 
While these approaches are promising, they suffer from at least one of three issues: (a) severe memory overhead, (b) high index construction time, and (c) low query throughput. 
This renders these methods practically unusable on larger arrays: For example, the state-of-art GPU-based approaches \LCA{} and \RTXRMQ{} exceed the memory capacity of an NVIDIA RTX 4090 GPU for input arrays of size $ \ge 2^{29}$. 
To tackle these problems, in this work, we present a new approach called \ours{} which is based on a hierarchical approach. \ours{} first constructs a hierarchy of range minimum summaries on top of the original array in a highly parallel fashion. For query answering, only the relevant portions of the hierarchy are then processed in an optimized massively-parallel scan operation. 
Additionally, \ours{} is hybrid in design, enabling the use of both ray tracing cores and CUDA cores across different levels of the hierarchy to handle queries.
Our experimental evaluation shows that \ours{} outperforms the state-of-the-art approaches in terms of query throughput especially for larger arrays while offering a significantly lower memory footprint and up to two orders-of-magnitude faster index construction. In particular, it achieves up to $\sim8\times$ higher throughput than \LCA, $\sim17\times$ higher throughput than \RTXRMQ{}, and up to $\sim4800\times$ higher throughput compared to an optimized CPU-based approach.
\end{abstract}

\maketitle

\ifdefempty{\vldbavailabilityurl}{}{
\vspace{.3cm}
\begingroup\small\noindent\raggedright\textbf{Artifact Availability:}\\
The source code, data, and/or other artifacts have been made available at \url{\vldbavailabilityurl}.
\endgroup
}

\section{Introduction}

Given an array $X$ of $n$ numbers, range minimum queries (RMQs) address the problem of finding the minimum value within arbitrary contiguous sub-arrays of~$X$. They are fundamental operations and can be used for finding lowest common ancestors in trees, computing longest common extensions of suffixes, or answering maximum-sum segment queries. They are featured in a wide range of application areas including biological sequence analysis, and document retrieval.
Consider, for example, the state-of-the-art long read aligner {\it Minimap2} \cite{PAPER_MINIMAP}, one of its key components, the chaining module, relies on solving RMQs and accounts for up to 35\% of the tool's total computational runtime \cite{PAPER_MINIMAP_STAT}, or state-of-the-art methods for distributed suffix array construction and querying \cite{flick2019distributed}.
Motivated by the importance of RMQs across such diverse range of applications, their acceleration has been explored using a variety of architectures such as CPUs \cite{PAPER_RMQ_APPLICATIONS, PAPER:FischerHeun, CPU, PAPER:HRMQ_extended, kowalski2018faster, FerraginaL25, russo2022range, alzamel2017answer}, GPUs \cite{PAPER:CTREE, PAPER:GPU:LCA_with_Euler, meneses2024accelerating}, and even QPUs \cite{wang2026quantum}.

\begin{figure}[htbp]
    \begin{subfigure}[t]{0.34\columnwidth}
        \includegraphics[width=\linewidth]{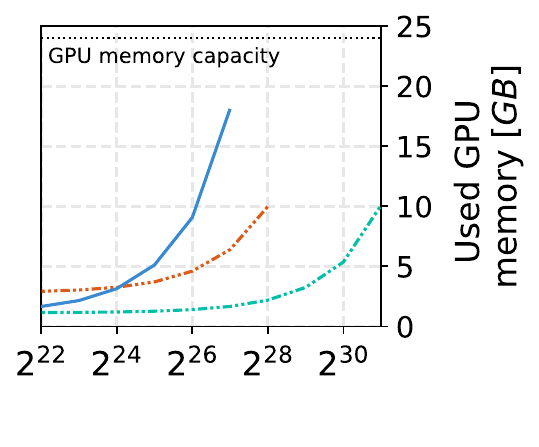}
        \caption{Memory footprint.}
        \label{fig:introduction_c:memory}
    \end{subfigure}%
    \begin{subfigure}[t]{0.34\columnwidth}
        \includegraphics[width=\linewidth]{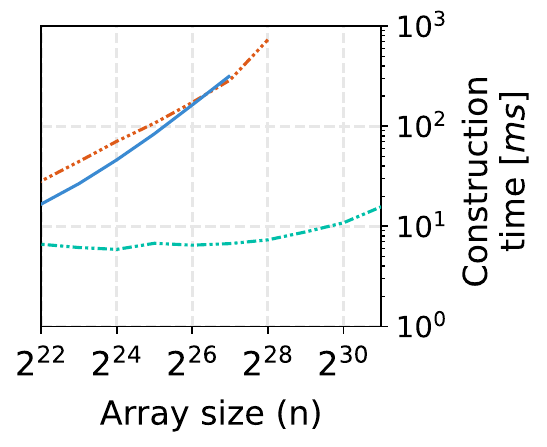}
        \caption{Construction time.}
        \label{fig:introduction_b:constr_time}
    \end{subfigure}%
    \begin{subfigure}[t]{0.34\columnwidth}
        \includegraphics[width=\linewidth]{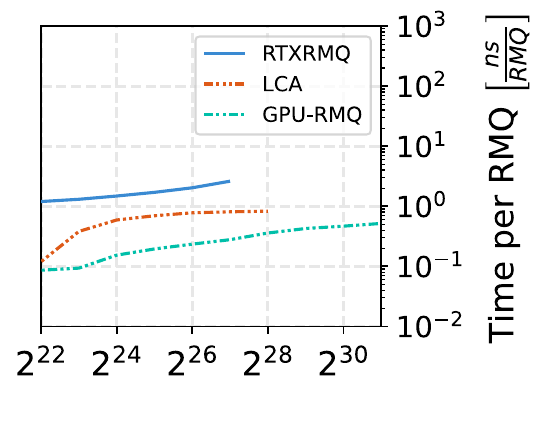}
         \caption{RMQ Throughput.}
        \label{fig:introduction_a:query_time}
    \end{subfigure}%
    \caption{\ours{} addresses important limitations of the prior GPU-based approaches \RTXRMQ{} and \LCA{}, resulting in (a) lower memory footprint, (b) significantly faster index construction times, and (c) higher throughput per RMQ (results shown for the mixed datasets (see Section \ref{sec:setup}) on an RTX 4090 GPU).}
    \label{fig:introduction}
\end{figure}

For static RMQs (i.e. batched range minima are computed without updating $X$), a data structure is typically built for the input array in a preprocessing step, which is then used to answer batches of RMQs. Prior GPU-based approaches rely on different types of tree data structures. Soman et al. \cite{PAPER:CTREE} were the first to implement RMQs on GPUs by means of a Cartesian tree data structure, while Polak et al. \cite{PAPER:GPU:LCA_with_Euler} proposed \LCA{} -- a method to compute Euler tours on trees that can be used for both lowest common ancestor queries and RMQs.  
Meneses et al. \cite{meneses2024accelerating} recently proposed \RTXRMQ{} -- a new method to leverage RT (ray tracing) cores to solve RMQs.
Their key idea is to represent all values of the input array as a geometric scene.
By launching multiple rays into this scene, RMQs can be solved in parallel by taking advantage of hardware-accelerated bounding volume hierarchy (BVH) traversal capabilities on GPUs. 

Unfortunately, all these approaches suffer from significant memory consumption to store the corresponding data structures in GPU memory. 
For example, the geometric scene constructed by \RTXRMQ{} requires at least one triangle per array element, which limits its applicability for large input arrays due to insufficient memory resources. In addition, traversing trees on GPUs can be inefficient due to irregular memory access schemes.
Figure \ref{fig:introduction}, which shows (a) memory footprint, (b) index construction time, and (c) RMQ throughput of the two state-of-the-art GPU approaches \LCA{} and \RTXRMQ{} on a NVIDIA RTX 4090 for various input array sizes, presents these problems: For an input array consisting of $n = 2^{29}$ elements, the memory usage of both \LCA{} and \RTXRMQ{} already exceeds the capacity of the utilized RTX 4090 GPU (24 GB). Further, both \LCA{} and \RTXRMQ{} suffer from a high construction time, which increases heavily with an increase in input size. Also, for larger input arrays that exceed the available cache, the time per RMQ heavily increases. 

In this work, we address these limitations by proposing an efficient hierarchical data structure and a corresponding processing scheme called \ours{}. 
On a high-level, \ours{} creates a hierarchy of range minimum summaries on top of the original array. To answer a batch of RMQs, \ours{} then scans and post-filters the relevant parts of the hierarchy via highly optimized hardware-conscious scan operations using cooperative thread groups. As we can see in Figure~\ref{fig:introduction}, this leads to a high RMQ throughput \textit{while} consuming only a moderate amount of auxiliary memory and offering a low construction time for the data structure --- enabling GPU-based processing of significantly larger arrays. Additionally, we designed \ours{} as a hybrid approach in which the upper level of the hierarchy can be processed by RT cores while lower levels are handled by CUDA cores, thereby combining different hardware capabilities of modern GPUs.

\subsection{Contributions and Structure of the Paper}

In summary, we make the following contributions in this work:
\begin{enumerate}
    \item We propose \ours{}, a highly efficient GPU-based acceleration structure for answering RMQs. On a high-level, \ours{} builds a hierarchy of range minimum summaries of varying granularity. For query answering, \ours{} identifies and scans the relevant parts of this hierarchy in a highly parallel fashion to compute the actual minimum. 
    \item \ours{} integrates a set of carefully designed optimizations that aim at exploiting the characteristics of the highly-parallel hardware. This includes vector loading and coalesced loading to speed up scanning, as well as multi-loading and warp-local queuing for assigning queries to hardware threads. As these optimization techniques are configurable, we experimentally determine their impact and configuration of choice.  
    \item We designed \ours{} from ground up as a hybrid acceleration structure, which supports the efficient usage of not only CUDA cores, but also RT cores. As such, the top-most level of the hierarchy can be expressed as a triangle scene in which hardware-accelerated ray-triangle intersection tests are performed instead of scanning. Again, we determine the effectiveness of such a hybrid approach experimentally. 
    \item In an extensive experimental evaluation, we show that \ours{} successfully addresses the three main disadvantages of the state-of-the-art methods \LCA{}, \RTXRMQ{}, and \HRMQ{}, namely severe memory overhead, high construction time, and low query throughput for a variety of workloads. As a result, \ours{} is able to handle significantly larger datasets than the baseline methods on the available GPU memory \textit{while} offering a superior query throughput for a large range of dataset sizes. More specifically, our GPU memory footprint is up to $\sim4\times$ smaller than \LCA{} and $\sim10\times$ smaller than \RTXRMQ. Our construction time is up to $\sim100\times$ faster than \LCA{}, $\sim50\times$ faster than \RTXRMQ{}, and $\sim2400\times$ faster than \HRMQ. Our throughput is up to $\sim8\times$ faster than \LCA{}, up to $\sim17\times$ faster than \RTXRMQ{} and up to $\sim4800\times$ faster than \HRMQ{}. 
\end{enumerate}

The rest of the paper is organized as follows.
After providing background in Section \ref{sec:background} and reviewing related work in Section \ref{sec:related}, we present the core design of \ours{} as well as the applied optimizations in Section \ref{sec:GPU-RMQ}. 
We then perform an extensive experimental evaluation in Section~\ref{sec:results}, which we kick off by identifying the optimal configuration for \ours{} for a variety of relevant array sizes. After that, we compare \ours{} against a set of state-of-the-art CPU and GPU baselines in terms of (i)~memory footprint, (ii)~construction time, and (iii)~RMQ throughput. Additionally, we (iv)~analyze the performance across different GPU generations.
 Finally, we conclude in Section~\ref{sec:conclusion}.

\section{Background}\label{sec:background}

Let us first discuss all necessary background on RMQs as well as on the hardware on which they will be executed. 

\subsection{Range Minimum Queries (RMQs)}

\begin{figure}[h!]
	\centering
	{\includegraphics[width=.95\linewidth, page=1, trim={0 12.3cm 17.5cm 5cm}, clip]{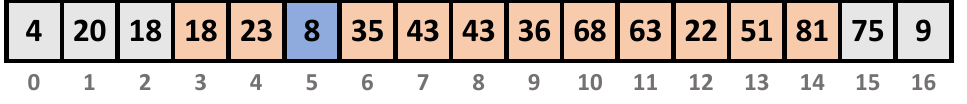}}
	\caption{Scanning the interval $(3, 14)$ in a 17-element input array for an RMQ. The query would return $\text{RMQ}_\text{value}(3,14) = 8$ or $\text{RMQ}_\text{index}(3, 14) = 5$ (blue highlight).}
	\label{fig:background:example_RMQ}
\end{figure}

We consider an array $X = [x_0, x_1, \dots, x_{n-1}]$ of $n$ numbers.
\textbf{Range Minimum Queries} (RMQs) answer queries consisting of pairs of positions $(l, r), \ 0 \leq l \leq r < n$, by either returning the minimal value within the range from  $l$ to $r$:
\[\text{RMQ}_\text{value}(l, r) = \underset{l \leq k \leq r}{\min} \ x_k\]
or by returning an index associated with a minimum:
\[\text{RMQ}_\text{index}(l, r) = \underset{l \leq k \leq r}{\arg\min} \ x_k.\]
An example for answering the RMQ for $(l, r)=(3, 14)$ on an array of 17~entries via a simple scan is shown in \autoref{fig:background:example_RMQ}.

Note that we focus on the answering of a batch of RMQs on a \textit{static} input array, as it is the typical scenario \cite{PAPER:FischerHeun, liu2020lower, FerraginaL25, meneses2024accelerating} for RMQs. This allows for the preprocessing of~$X$ to build a data structure upfront that can be used to speed up the RMQs. Thus, approaches can be categorized by a space-time trade-off; i.e., the time complexity for solving a single RMQ and the space complexity for storing the additional data structure. The two extreme cases are: 
\begin{itemize}
    \item {\bf Full preprocess:} A data structure of size $\mathcal{O}(n^2)$ is built that stores the results for all possible ranges. Each RMQ  is then solved in constant time $\mathcal{O}(1)$ by a simple lookup.
    \item {\bf Full scan:} No preprocessing is done and each RMQ is performed by scanning all elements within the corresponding range in time $\mathcal{O}(n)$.   
\end{itemize}
While the former case requires too much memory, the latter is too slow for practical applications that frequently have to compute the minimum for large ranges. As a consequence, existing approaches focus on finding solutions between these two extremes.  


\subsection{Modern GPU Architectures}

\begin{figure}
    \centering
    \includegraphics[width=.95\linewidth, page=11, trim={0 11cm 14.5cm 0}, clip]{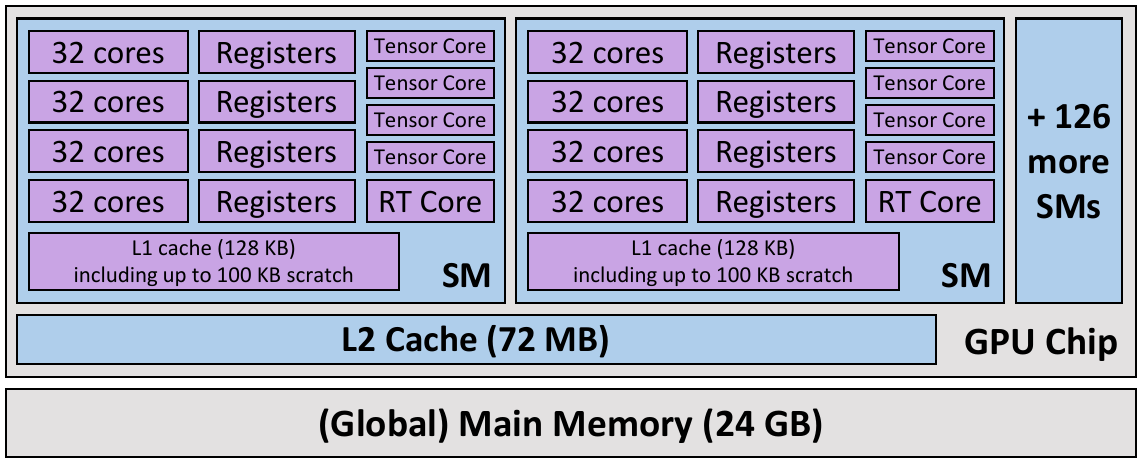}
    \caption{Architecture of the NVIDIA RTX 4090.}
    \label{fig:GPU}
\end{figure}

As our main testbed, we use a recent high-end consumer GPUs: NVIDIA RTX~4090 (Ada architecture) with 24GB of main memory. The architecture is visualized in Fig. \ref{fig:GPU}. Processing on the 4090 is distributed across 128 streaming multiprocessors (SMs) containing 4 vector units of 32 cores as well as one RT core each. Typically, one unit of work is assigned to one thread. When running a multi-threaded task, the GPU distributes small batches of threads (thread blocks) across SMs, which then schedule even smaller batches of 32 threads (warps) onto the 32-core vector units. Each SM has its own L1 cache (128 KB), which can be partially (up to 100 KB) reconfigured to serve as fast user-programmable scratch memory, allowing for fast data movement and storage within a thread block. Similarly, each SM contains 256 KB worth of 32-bit registers, which can be cross-referenced by threads within a warp using special instructions (warp intrinsics) to exchange data on-the-fly (warp shuffling). Additionally, the GPU has a global L2 cache (100 MB). To ensure scalability, direct communication across SMs is restricted to communication though the GPU global main memory.
Note that in Section~\ref{ssec:different_gpu_generations}, we will also investigate the performance for two other GPU generations, namely (1)~an NVIDIA~RTX~PRO~6000 (Blackwell architecture) with 96GB of global memory and 188~SMs and (2)~an NVIDIA RTX 3090 (Ampere architecture) with 24GB of global memory and 82~SMs. 

\begin{figure}
	\centering
	\includegraphics[width=.95\linewidth]{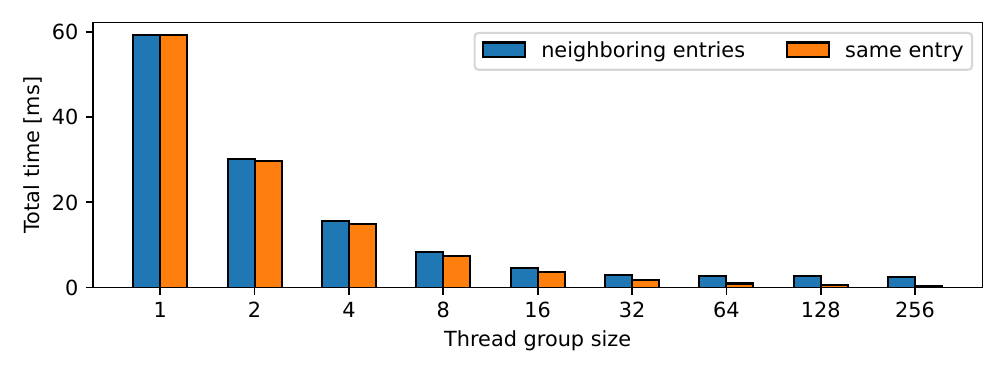}
	\caption{Execution time for a batch of coalesced accesses.}
	\label{fig:micro:cg}
\end{figure}

A notable feature of GPU memory is \textit{access coalescing}:
When multiple neighboring threads within a warp load neighboring entries at the same time, the GPU performs a single, larger memory access instead.
To demonstrate this effect with a micro-benchmark, we set up an array containing $2^{28}$ entries of 32 bits each, populated with random data.
Then, a total of $2^{25}$ GPU threads are launched, organized into groups of $2^m$ adjacent threads, yielding $2^{25-m}$ non-overlapping groups.
Each group carries out $1024$ lookup iterations: At every iteration, the group selects a random array index $p$ to read from, which simulates divergent access patterns for widely separated reads.
Within a given group, thread $i$ reads from index $p+i$, meaning the group collectively accesses $2^m$ consecutive but distinct entries.
The overall number of load operations stays the same across configurations, only the access pattern varies.
Figure~\ref{fig:micro:cg} shows the total runtime of these lookup iterations across different group sizes.
The results reveal that doubling the group size cuts execution time roughly in half, up to a size of 16.
Increasing to 32 threads per group still helps, though to a lesser degree, while going beyond 32 provides no additional speedup.
For reference, Figure~\ref{fig:micro:cg} also includes timings for the case where every thread in a group reads only the single entry at position $p$.
The performance gain is even more pronounced there, as the hardware can cache and reuse that one value.


\section{Related Work}
\label{sec:related}

Before presenting \ours{}, we discuss the related work on GPU-based data structures in general, and on RMQ approaches in particular. 

\subsection{GPU-based Data Structures}

In recent years, various well-known CPU-based data structures have been transitioned successfully to GPUs. These works share with ours that they also base their design around the specifics of the hardware architecture, in particular the extremely high parallelism offered by the card. We believe that fundamental optimization techniques employed by our work, such as coalesced scanning and warp-local queuing, can also be utilized to enhance other scan-heavy general-purpose data structures.

For instance, Awad et al.~\cite{awad:btree} proposed a flexible concurrent B-tree implementation for GPUs, supporting point- and range-lookups as well as an efficient on-card bulk-loading mechanism. Apart from parallelizing across lookups, it further exploits the high parallelism of the GPU by having groups of 16 threads inspect the content of a node concurrently within a single lookup in order to determine the next node respectively the key of interest in the leaves. Unfortunately, it is currently limited to 32-bit keys and values. 
In a similar style, Ashkiani et al.~\cite{ashkiani:lsm} propose a GPU-based LMS-tree implementation, which focuses on supporting efficient updates. It utilizes the parallelism of the GPU to sort and merge the individual levels of the LSM-tree.
If range-lookups are not required and point-lookups dominate, hash tables are to be preferred. Again, Ashkiani et al.~\cite{ashkiani:slabhash} proposed in form of slab-hash an update-focused implementation for that. Further, Jünger et al.~\cite{juenger:warpcore} proposed Warpcore, a highly efficient lookup-focused open-addressing hash table, which combines warp-wide linear probing with outer-layer double hashing to minimize clustering.   
In the presence of static datasets, Henneberg et al.~\cite{henneberg:eytzinger} question the need for sophisticated data structures altogether, and proposed to use an highly optimized variant of $k$-ary search on a sorted array instead. To efficiently scan for qualifying entries during range-lookups, they also effectively utilize coalesced scanning, where neighboring threads handle the scanning neighboring memory regions. 
Further successful examples of GPU-based general-purpose data structures include Bloom filters \cite{juenger:bloomfilter}, skiplists \cite{moscovici:skiplist}, and  FIFO-stacks \cite{south:stack}.

While the original application of RT cores is the production of real-time photorealistic graphics, recent work has examined their utility for non-rendering problems. 
A typical approach to adapt a considered task to RT cores consists of two phases: (i) encoding the input data as 3D geometric objects in a scene and building the corresponding BVH to serve as an auxiliary index structure; (ii) casting a large number of rays to perform lookups; i.e., if a ray intersects with an object, the lookup returns an identifier associated with the object as a search result. Examples of applications following this approach include scans \cite{lv2024rtscan}, database queries \cite{PAPER_RTIndeX, henneberg2025more, shi2025raydb}, nearest neighbor search \cite{related_work:NearestNeighbor2, shi2025raydb}, or spatial joins \cite{geng2024rayjoin, geng2025librts}. To our knowledge we are the first to investigate the collaborative usage of CUDA cores and RT cores for non-rendering problems with a single data structure. 

\subsection{State-of-the-Art RMQ Approaches}
\label{ssec:existing_approaches}

One option to answering RMQs is to reduce it to the problem of finding the LCAs in the Cartesian tree of the input array. This approach offers a space complexity of $\mathcal{O}(n \cdot \log(n))$ at $\mathcal{O}(\log(n))$ query time. The Inlabel algorithm~\cite{PAPER:Inlabel_algorithm} improved this by achieving constant-time LCA queries with linear space consumption.  

\HRMQ{}. Ferrada et al. \cite{PAPER:HRMQ_extended} introduced a CPU-based approach for answering RMQs, building upon the design of Fischer and Heun~\cite{CPU}. Instead of the previously used Depth-First Unary Degree Sequence (DFUDS), they use a balanced parentheses representation. Their algorithm keeps the asymptotically optimal size of $2n + o(n)$ and maintains constant query time. While being considered one of the fastest CPU-based implementation, its performance is limited by the low degree of parallelism available on multi-core CPUs in comparison to massively-parallel GPUs architectures. 





\LCA{}. Soman et al. \cite{PAPER:CTREE} were the first to solve RMQs on GPUs by introducing a succinct representation based on DFUDS designed for efficient usage on the GPU.
They report a speedup ranging from $25\times$ to $35\times$ compared to a CPU-based method provided by Fischer and Heun~\cite{PAPER_RMQ_APPLICATIONS}.
By employing the Euler tour technique, Polak et al. \cite{PAPER:GPU:LCA_with_Euler} proposed \LCA{} which adapts the Inlabel algorithm for GPUs.
Their evaluation reports speedups ranging from $22\times$ to $55\times$ compared to a single-threaded CPU implementation of the Inlabel algorithm. 
However, the algorithm has to materialize (a)~the order of nodes visited on the Euler tour over its Cartesian tree, (b)~the depth of each node visited, as well as (c)~the first occurrence of each node on the tour. This leads to high memory consumption; e.g. e.g. for $n=2^{29}$ it exceeds the memory capacity of an RTX4090 GPU (24GB).

\begin{figure}
    \centering
    \includegraphics[scale=0.12]{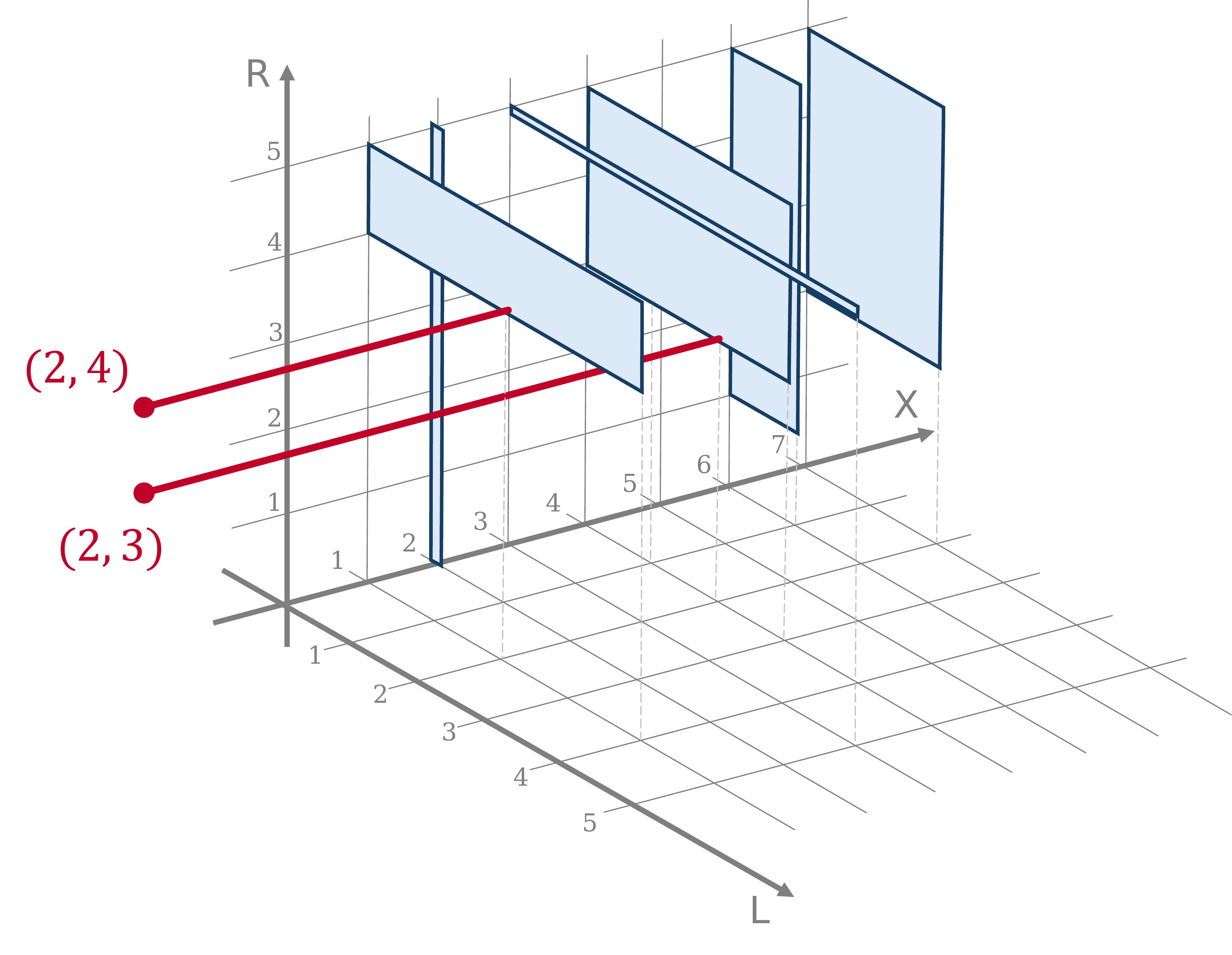}
    \caption{\RTXRMQ{} solving two RMQs on $X = [2, 6, 7, 4, 1, 3]$ with RT cores.}
    \label{fig:3D_scene}
\end{figure}

\RTXRMQ{}. Recently, \RTXRMQ{} \cite{meneses2024accelerating} has been proposed to leverage hardware-accelerated BVH of RT cores on modern RTX video cards for solving RMQs. It launches one ray for each RMQ. The ray for $\text{RMQ}(l, r)$ is launched from the coordinate $(- \infty, l, r)$ in direction $(1, 0, 0)$. It can only intersect with geometric primitives corresponding to array elements $x_i$ with $l \leq i \leq r$. To achieve this one geometric primitive for each array element is created in a preprocessing step and positioned along the X-axis according to the element’s value. By launching rays along the X-axis, the first intersection of the ray will identify the triangle corresponding to the range minimum of the array (illustrated in Fig. \ref{fig:3D_scene}). 
A major limitation of \RTXRMQ{} is its excessive memory consumption since its data structure consists of a large number of geometric primitives stored in a BVH; e.g. for $n=2^{28}$ it already exceeds the memory capacity of an RTX4090 GPU (24GB). Additionally, the construction of the BVH is an expensive operation.

In summary, there already exist a number of RMQ algorithms for both CPU and GPU.
Although CPU-based approaches can be asymptotically efficient they often suffer from slow runtimes due to limited parallelism. While GPU-based methods expose massive parallelism they operate on tree data structures, posing challenges for efficient global memory access. In addition, the GPU global memory can be a scarce resource and is typically much smaller than CPU main memory. Thus, memory consumption of the auxiliary data structure becomes a bottleneck. This establishes the need for a new approach that can exploit the fast speed of GPUs for RMQs while consuming significantly less memory than prior approaches.


\section{GPU-RMQ}\label{sec:GPU-RMQ}

We first introduce the core \ours{} data structure and its algorithms, and then describe the GPU-specific optimizations we have applied to improve its performance.

\subsection{Core Data Structure}
\label{ssec:core}

In theory, every RMQ can be answered by sequentially traversing the query range of the input array and computing the minimum.
We visualized this in our introductory example in Figure~\ref{fig:background:example_RMQ}, where 12 entries have to be traversed to find the minimum between positions 3 and 14 of the input array.
However, while GPUs excel at sequential array traversals (aka scans), the cost for analyzing larger ranges becomes prohibitively high.  

\begin{figure}
	\centering
	{\includegraphics[width=.95\linewidth, page=2, trim={0 11.6cm 17.5cm 5cm}, clip]{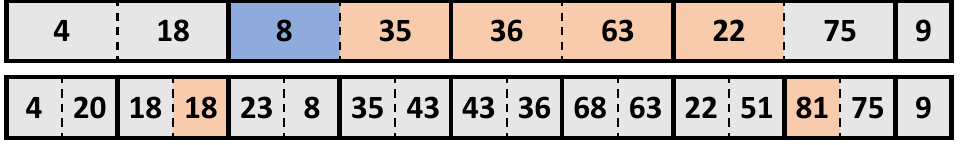}}
	\caption{Running $\text{RMQ}_\text{value}(3, 14)$ on a two-layer minima hierarchy.}
	\label{fig:alg:two_levels}
\end{figure}

\begin{figure}
	\centering
	{\includegraphics[width=.95\linewidth, page=3, trim={0 10.3cm 17.5cm 5cm}, clip]{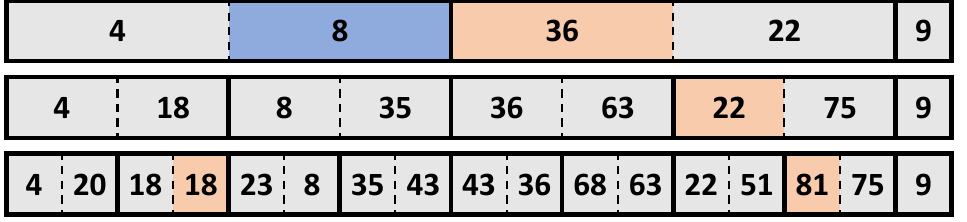}}
	\caption{Running $\text{RMQ}_\text{value}(3, 14)$ on a three-layer minima hierarchy.}
	\label{fig:alg:three_levels}
\end{figure}

To drastically reduce this cost, we can pre-compute the minima of fixed-size chunks of size~$c$ (typically a power of two) and store them in a separate array as an additional layer.
This reduces the problem to a coarse-granular RMQ in the significantly smaller minima array, followed by searching the minimum in the two chunks on the bottom layer that are not fully covered by the query.
We illustrate this in Figure~\ref{fig:alg:two_levels}, where $c = 2$ neighboring entries form a chunk (separated by dashed lines).
Here, the query can be answered by scanning two entries in the bottom layer, and five pre-computed minima in the new layer, almost halving the amount of scanned entries.
Of course, this pre-computation step can be applied recursively, producing a multi-level hierarchy of minima arrays.
A three-layer hierarchy is shown in Figure~\ref{fig:alg:three_levels}, where we only have to traverse five entries in total to answer the RMQ.
In theory, building a four-layer hierarchy would be possible for our example, but it would not be beneficial to our query since the query does not fully contain any third-layer chunk.

This approach is efficient in terms of both construction and storage overhead:
Generating the upper layers can be parallelized efficiently on a GPU by constructing each auxiliary layer of the hierarchy from bottom to top in parallel, where a group of adjacent threads reduces a chunk of $c$~adjacent entries to a single minimum summary. The resulting size of the hierarchy is only a small fraction of the input array's memory footprint, as the number of additional elements $E$ in the hierarchy is bounded by
\begin{align*}
    E &\le \sum_{i=1}^{\log_{c}(n)} \frac{n}{c^i} 
     \le \frac{n}{c}  \, \sum_{i=0}^{\infty} \left(\frac{1}{c}\right)^i
     = \frac{n}{c} \, \frac{1}{1 - \frac{1}{c}} = \frac{n}{c} \, \frac{c}{c-1} = \frac{n}{c - 1}.
\end{align*}

To further reduce allocation complexity, we store all pre-computed layers in a single, contiguous buffer.
Answering a query only requires a sequential scan of the topmost qualifying layer, followed by a few random but cache-aligned chunk accesses on the lower layers.
The shape of the structure is determined by two parameters:
The chunk size $c$, and a threshold $t$, which defines the maximum number of chunks allowed on the topmost layer.
Theoretically, we have to scan at most $ct + 2c\log_{c}(n) \in \mathcal{O}(\log(n))$ entries to answer a query.
In Section~\ref{ssec:tuning_parameters}, we will identify the best choices for both parameters~$c$~and~$t$. 

\subsection{Scan: Vector Loading vs Coalesced Loading}
\label{ssec:vector_vs_coalesced_loading}

Apart from the parameter choices, it is critical how to implement the scan. \ours{} supports the following two options in maximizing scan throughput on a GPU:

\textbf{(1)~Vector loading}. Instead of sequentially accessing single entries, a thread can issue an explicit \textit{vector load} instruction, which loads multiple neighboring entries from memory into the thread's registers, but costs the same as a single-entry access.
Vector loads require the data to be aligned to the vector size, and are only available for up to four entries.
Applied to our RMQ problem, scanning performance can be improved by fixing the chunk size to a multiple of the vector size, and access the chunk using a sequence of vector loads.
After each vector load, we sequentially compute the minimum across the vector entries (now stored in registers), ignoring the entries that do not fall within the range of positions covered by the RMQ.
An individual vector load can be skipped altogether if none of the vector entries qualify for the query.

\textbf{(2)~Coalesced loading}. Alternatively, we can use a multi-threaded strategy:
We assign a group of $g \le 32$ neighboring threads to the same RMQ, and use them to load neighboring entries within the same chunk.
This results in coalesced memory accesses, greatly increasing memory throughput.
We re-use this group across all lower-layer chunks, and also for the top-layer scan.
A thread skips the load operation if its entry is outside the RMQ bounds.
In contrast to the vector load strategy, each thread now independently tracks the minimum of the (qualifying) entries it accesses.
After the last traversal step, we compute the minimum across the thread group using warp shuffles (or by atomic aggregation when shuffling is unavailable).

\begin{lstlisting}[language=C++, caption={Coalesced scan algorithm.}, label={lst:scan}, breaklines=true]
float scan(group, l, r, array, m) {
  uint my_rank = group.thread_rank();
  // align accesses to multiples of g
  uint offset = l - l % g;
  for (; offset < r; offset += g) {
    uint my_offset = offset + my_rank;
    if (l <= my_offset && my_offset < r) {
      m = min(m, array[my_offset]);
    }
  }
  return m;
}
\end{lstlisting}

Listing~\ref{lst:scan} shows the code for scanning a section of an array using coalesced loading and computing the thread-local minima.
Listing~\ref{lst:hierarchy} shows the full RMQ subroutine, which determines the bounds for the individual scans up the hierarchy, and performs the intra-group reduction.
Note that both code snippets assume some information to be globally available, such as the chunk size \texttt{c}, thread group size \texttt{g}, and the two arrays \texttt{base\_array} for the bottommost layer as well as \texttt{upper\_array} for the remaining layers.

\begin{lstlisting}[language=C++, caption={Scanning the hierarchy.}, label={lst:hierarchy}, breaklines=true]
float rmq(group, l, r) {
  uint level = 0;
  float m = +INFINITY;
  r += 1;

  for (; level < num_levels - 1; ++level) {
    if (r - l <= 2 * c) {
        break;
    }

    // next multiple of c larger or equal to l
    uint32_t next_l = (l + c - 1) - (l + c - 1) % c;
    // next multiple of c smaller or equal to r
    uint32_t prev_r = r - r % c;

    // layer zero is the original array
    auto array = level == 0
      ? base_array
      : upper_array + level_offset[level];
    // scan left and right chunks
    m = scan(group, l, next_l, array, m);
    m = scan(group, prev_r, r, array, m);

    l = next_l / c;
    r = prev_r / c;
  }

  auto array = level == 0
    ? base_array
    : upper_array + level_offset[level];
  // scan last layer
  m = scan(group, l, r, array, m);

  return cg::reduce(group, m, cg::less<float>());
}
\end{lstlisting}

\begin{figure}
	\centering
	{\includegraphics[width=.95\linewidth, page=5, trim={0 14cm 17.5cm 3cm}, clip]{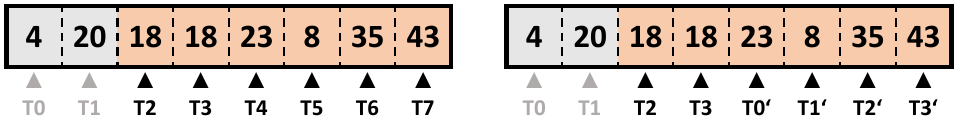}}
	\caption{Scanning a chunk with 6 entries in the RMQ range using larger (left) and smaller (right) thread groups.}
	\label{fig:alg:groupsize}
\end{figure}

\begin{figure}
	\centering
	{\includegraphics[width=.95\linewidth, page=5, trim={0 16.8cm 17.5cm 0cm}, clip]{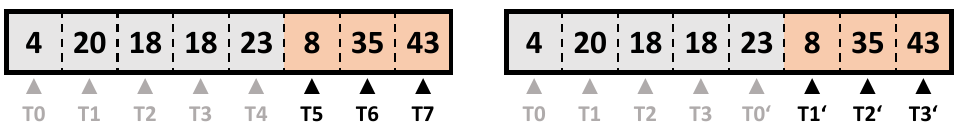}}
	\caption{Scanning a chunk with 3 entries in the RMQ range using larger (left) and smaller (right) thread groups.}
	\label{fig:alg:groupsize2}
\end{figure}

The group size $g$ can be chosen independently from the chunk size $c$.
Choosing a large group size increases the memory access efficiency of a single RMQ, but reduces the number of RMQs that we can process concurrently, as only a fixed number of threads is active at any given time.
In addition, choosing $g$ close to $c$ can lead to poor thread usage:
Figure~\ref{fig:alg:groupsize} shows an RMQ with $6$~entries in its range in a chunk of size $c=8$ being answered using a single access with group size $g=8$ (left) or two accesses with $g=4$ (right).
Each thread in the group only loads and processes an entry it is assigned to if the entry falls into the range of the RMQ.
In the figure, these qualifying entries are highlighted in orange, while non-qualifying entries are shown with a gray backdrop.
When $g=8$, threads 0 and 1 stay idle during the scan because they are assigned to non-qualifying entries, and the remaining six threads in the group process the remaining entries.
With $g=4$, there are only four active threads, and they perform the scan in two steps.
In the first step, threads 0 and 1 stay idle again.
However, in the second step, all four threads can do useful work, since all entries in the second half of the chunk fall into the range of the query.
It is evident that choosing $g=4$ leads to higher thread utilization in this example.
The effect is even more exaggerated in Figure~\ref{fig:alg:groupsize2}:
Choosing $g=8$ results in five idle threads.
In contrast, when $g=4$, the first of the two loads is skipped altogether (since no entries in the first half fall into the range), and only one thread remains idle during the second step.
In addition to higher thread utilization, choosing a smaller $g$ also saved memory bandwidth in this example.
Overall, setting~$g$ is non-trivial, which is why we evaluate it experimentally in Section~\ref{ssec:tuning_parameters}.

\subsection{Query Assignment: Multi-load vs Warp-local Queuing (WLQ)}\label{sec:query_assignment}

Remember that our main workload consist of batch-solving $m$~RMQs concurrently, where each RMQ $j$ is defined by its two bounds, $l_j$ and $r_j$.
Our previously discussed scan mechanisms require all participating threads of a group to work on the same RMQ. There are two strategies for achieving this:

\textbf{(1)~Multi-load}.
We spawn $g \cdot m$ threads in total.
The thread with index $tid$ is assigned to the query at position $j = \left\lfloor\sfrac{tid}{g}\right\rfloor$ within the batch, and therefore acquires $l_j$ and $r_j$ from main memory.
This approach is easy to implement and does not require communication between threads.

\begin{figure*}
	\centering
	\includegraphics[scale=0.102]{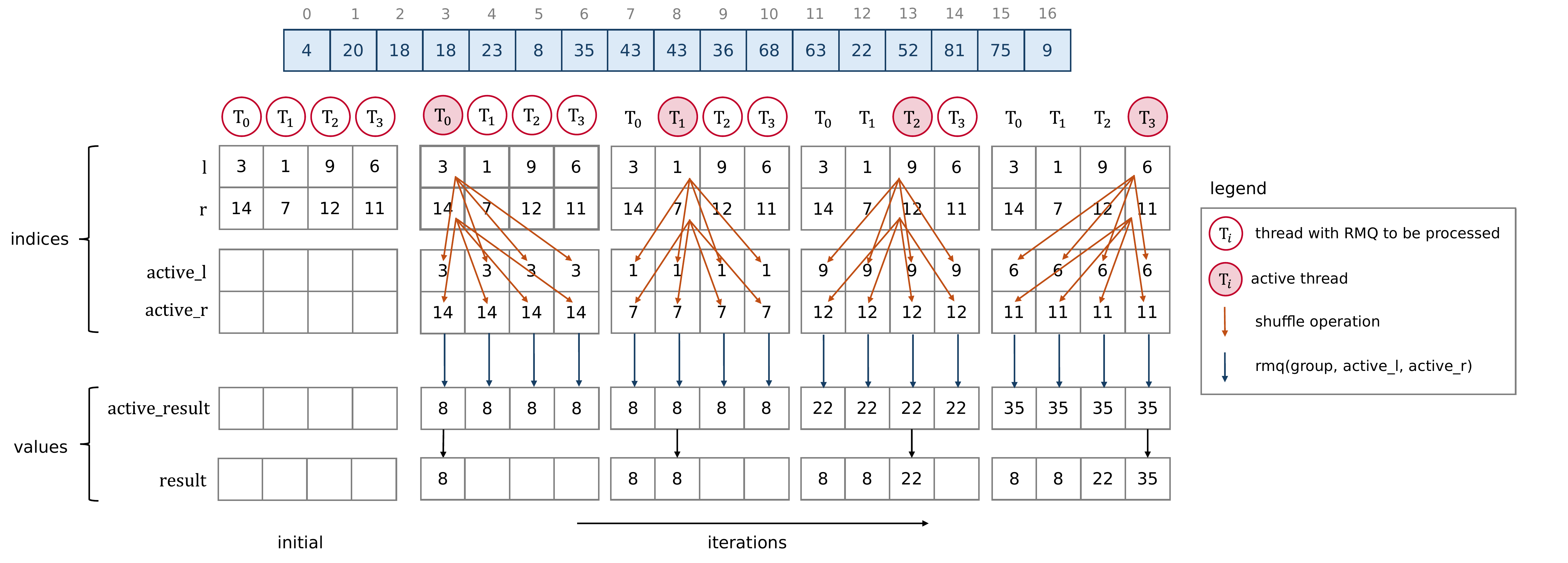}
    \caption{Example of WLQ in \ours{} with group size $4$. Each thread in the group has a distinct RMQ. To process them cooperatively, one thread is designated as active and shuffles its RMQ indices to the remaining threads. Each thread then calls the function \textit{rmq} (see \autoref{lst:hierarchy}), which returns the group-wide minimum computed with coalesced memory accesses. The active thread stores the final result, after which the next thread becomes active to process its RMQ.}
	\label{fig:method:pipeline}
\end{figure*}

\textbf{(2)~Warp-local queuing (WLQ)}.
We spawn only $m$ threads, rounded up to the next multiple of $g$.
The thread with index $tid$ loads $l_{tid}$ and $r_{tid}$.
Each group now performs $g$ many rounds:
For the first round, the first thread in the group uses warp shuffles to scatter its local values for $l$ and $r$ across the group.
The threads then proceed with processing the query together, just like we described in the previous section.
Finally, the computed result is sent back to the first thread.
In the second round, the second thread in the group scatters its RMQ to the other threads, and receives the result, and so on.
This essentially repurposes some of the threads' registers as a queue for the queries and results.
In the end, all threads write their results to some output buffer.
We provide a reference implementation for warp-local queuing in Listing~\ref{lst:queue} and illustrate the process in Figure~\ref{fig:method:pipeline}.

Assuming $g \le 16$, Strategy (1) performs $g \cdot m$ memory accesses, which the hardware coalesces into only $m$ actual transfers, while strategy (2) performs $m$ accesses, which the hardware coalesces into $\sfrac{m}{g}$ many.
In summary, Strategy (2) has noticeable better memory efficiency at the cost of increased register usage and requiring cross-thread communication.

\begin{lstlisting}[language=C++, caption={Warp-local queuing}, label={lst:queue}, breaklines=true]
uint my_id = my_thread_id();
auto group = my_thread_group(my_id);
uint my_rank = group.thread_rank();

float result;
uint l, r;
bool find;

if (my_id < num_queries) {
  l = queries_l[my_id];
  r = queries_r[my_id];
  find = true;
}

auto queue = group.ballot(find);

while (queue) {
  auto active_rank = __ffs(queue) - 1;
  auto active_l = group.shfl(l, active_rank);
  auto active_r = group.shfl(r, active_rank);

  auto active_result = rmq(group, active_l, active_r);
  if (active_rank == my_rank) {
    result = active_result;
    find = false;
  }
  queue = group.ballot(find);
}

if (my_id < num_queries) {
  result_buffer[my_id] = result;
}
\end{lstlisting}

\subsection{Returning the Index}

So far, we have only discussed how to answer $\text{RMQ}_\text{value}$.
Solving $\text{RMQ}_\text{index}$, which returns the index of the minimum, requires two small modifications:
(1) When building the minima hierarchy, in addition to the minimum itself, we also store the position of the minimum in the original array.
(2) When traversing the minima hierarchy, we return the stored position.
If the minimum occurs multiple times between $l$ and $r$, one can implement additional logic to guarantee we always return the rightmost or leftmost position.

\subsection{Using Ray Tracing in the Hierarchy}

Note that the scan on the topmost layer in \ours{} corresponds to a regular RMQ, albeit on a summarized version of the original dataset.
We can therefore introduce another dimension to our data structure:
Following the design of \RTXRMQ{}, we can replace the topmost layer with a triangle scene, and use the hardware ray tracing pipeline to answer the top-layer RMQ, while still applying our optimized scan strategy to the lower layers.

\begin{figure}[h!]
	\centering
	{\includegraphics[width=.95\linewidth, page=4, trim={0 9.5cm 17.5cm 5cm}, clip]{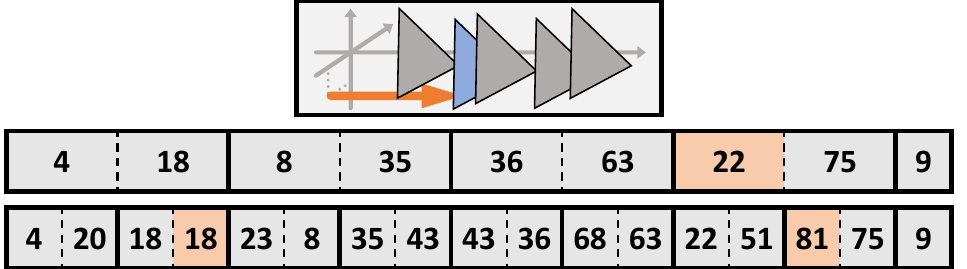}}
	\caption{Three-layer minima hierarchy with the top layer replaced by a triangle scene.}
	\label{fig:alg:three_levels_rt}
\end{figure}

We visualize this in Figure~\ref{fig:alg:three_levels_rt}, where the five triangles on the topmost layer replace the five array entries, and a ray-triangle intersection test replaces the scan.
This has the following theoretical implications:

(1) We can choose a larger value for the reduction cutoff threshold~$t$ because the complexity of the ray tracing job scales better with input size than a linear scan of the topmost layer, resulting in fewer overall layers.

(2) Since RT cores and CUDA cores are physically separate resources, the searches in the topmost layer and the remaining layers can be run concurrently within the same SM. 
To the best of our knowledge we are the first to exploit the combined usage of these different resources for a non-rendering problem with a single data structure. 

However, this combination also requires addressing a number of technical challenges.
In particular, the hardware ray tracing pipeline can only be accessed using the OptiX API, which is similar to CUDA, but introduces some restrictions to GPU programs.
Most significantly, warp shuffling is not available in OptiX, forcing us to use less efficient implementations for intra-group minimum reduction (Section~\ref{ssec:vector_vs_coalesced_loading}) and query assignment (Section~\ref{sec:query_assignment}).

Our OptiX implementation first triggers the ray tracing job, then performs the linear scan down the hierarchy.
Since there is no explicit dependency between the two parts, and GPUs heavily interleave warp execution anyway, this leaves plenty of leeway for the hardware to run ray tracing and scanning in parallel.

\section{Experimental Evaluation}\label{sec:results}

In the following, we first empirically identify the best parameter configuration for \ours{} (Section~\ref{ssec:tuning_parameters}), as well as the different options regarding query assignment and the usage of ray tracing (Section~\ref{ssec:evaluating_optimizations}). Then, we compare \ours{} against a set of state-of-the-art baseline methods in terms of memory footprint (Section~\ref{ssec:memory_footprint}), construction time (Section~\ref{ssec:construction_time}), and query time (Section~\ref{ssec:query_time}). Additionally, we analyze the behavior of \ours{} across different GPU architectures (Section~\ref{ssec:different_gpu_generations}). 

\subsection{Experimental Setup}\label{sec:setup}

As described earlier, we run the main line of experiments on an NVIDIA GeForce RTX 4090 (Ada) with 24GB of GPU memory, and use an AMD Ryzen Threadripper 3990X with 64 cores and 128 threads for GPU-CPU comparisons. 
Furthermore, the evaluation is conducted using CUDA 12.9.1 and OptiX 8.

To evaluate query performance, we test batches of three different range sizes in the following evaluation, namely large, medium, and small ranges, to study how much the range size impacts the performance of each method. These range sizes are defined as follows:
\begin{itemize}
    \item \textbf{Large ranges}: Uniformly distributed in $[1, n]$ with a mean of $\approx \frac{n}{2}$.
    \item \textbf{Medium ranges}: Log-Normal distributed with mean $\mu = \log(n^{0.6})$ and standard deviation $\sigma = 0.3$.
    \item \textbf{Small ranges}: Log-Normal distributed with $\mu = \log(n^{0.3})$ and $\sigma = 0.3$.
    \item \textbf{Mixed ranges}: Randomly sampled from the three previous types (large, medium, small) with equal probability.
\end{itemize}
The left border $l$ of a range is always drawn uniformly from $[0, n - s]$, where $s$ is the randomly chosen range size. 

To saturate the GPU, we fire a batch of $2^{26}$ RMQs in total unless mentioned otherwise.
The input array consists of independently sampled floating-point values from a uniform distribution on $[0, 1]$.
The queries are sent to the GPU as a buffer of $(l, r)$ 32-bit integer pairs.
We switch to 64-bit integers once the array size reaches $2^{31}$, to ensure every position in the array can be represented.

\subsection{Baseline Methods}

We evaluate our \ours{} against the following baselines, which have been described in detail in Section~\ref{ssec:existing_approaches}:

\exhaustive{}: A naive CUDA-based GPU implementation where each thread processes one RMQ using a full scan of the corresponding values in the input array.

\LCA{}: A state-of-the-art GPU-based approach for solving LCA queries by Polak et al.~\cite{PAPER:GPU:LCA_with_Euler}, where \RTXRMQauthors added the reduction from RMQ to LCA instances\footnote{The original implementation of \LCA{} and the adapted one can be found at \url{https://github.com/stobis/euler-meets-cuda} and \url{https://github.com/temporal-hpc/euler-meets-cuda-rmq}, respectively.}.

\RTXRMQ{}: The baseline that leverages RT cores exclusively to solve RMQs. \footnote{The implementation of \RTXRMQ{} is available at \url{https://github.com/temporal-hpc/rtxrmq}.}.
We tested an OptiX 8-adapted version of \RTXRMQ\@.
\RTXRMQ{} partitions the dataset with a user-configurable block size.
We tested a large number of block sizes for each value of $n$ and always report the result with the highest throughput.
Thus, the numbers serve as an upper bound for \RTXRMQ{}'s performance.

\HRMQ{}: A CPU-based approach based on the paper \textit{Improved Range Minimum Queries}~\cite{CPU}\footnote{The code of HRMQ can be found here: \url{https://github.com/hferrada/rmq}.}.
We use the OpenMP-parallelized version by \RTXRMQauthors and evaluate it with 128~CPU threads.

\subsection{Tuning Parameters of \ours{}}
\label{ssec:tuning_parameters}

As explained in \autoref{sec:GPU-RMQ}, \ours{} supports both vector loading and coalesced loading, which we differentiate as \oursVector{} and \oursCoalesced{} in the following.
As both variations provide several tunable parameters, namely the group size~$g$, the chunk size~$c$, and the build cutoff threshold~$t$, we will in the following empirically identify the best configuration for different array sizes~$n$. 

\begin{figure}[h!]
    \centering
    \includegraphics[width=\columnwidth]{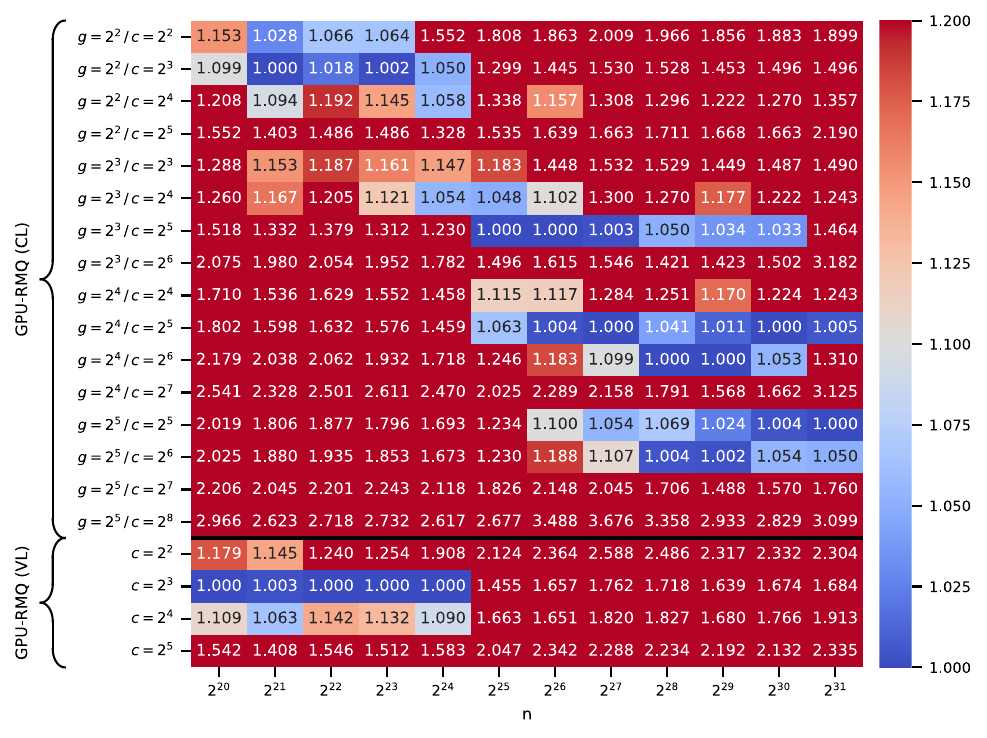}
    \caption{Query time comparison using different group sizes ($g$) and chunk sizes ($c$) on mixed ranges for \oursVector{} and \oursCoalesced. For each $n$, the values are normalized relative to the best runtime.}
    \label{fig:evaluation:plot:hyperparameter:heatmap}
\end{figure}

\begin{figure*}[h!]
    \centering

    \includegraphics[width=0.9\textwidth]{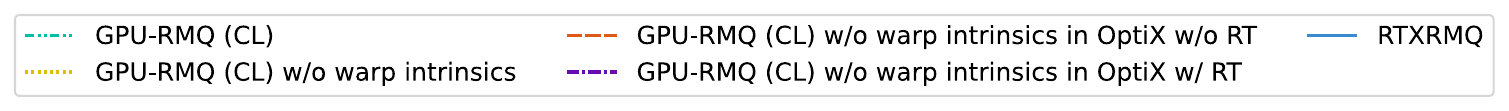}

    \begin{subfigure}[t]{0.31\textwidth}
        \includegraphics[width=\linewidth, clip]{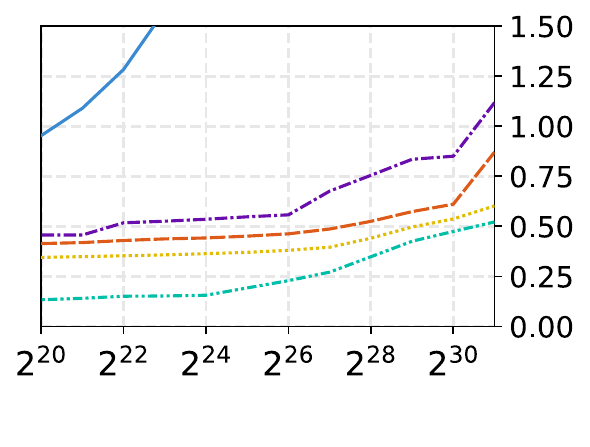}
        \caption{Large (l, r) range.}
        \label{fig:generations:variations:-1}
    \end{subfigure}%
    \hfill
    \begin{subfigure}[t]{0.31\textwidth}
        \includegraphics[width=\linewidth, clip]{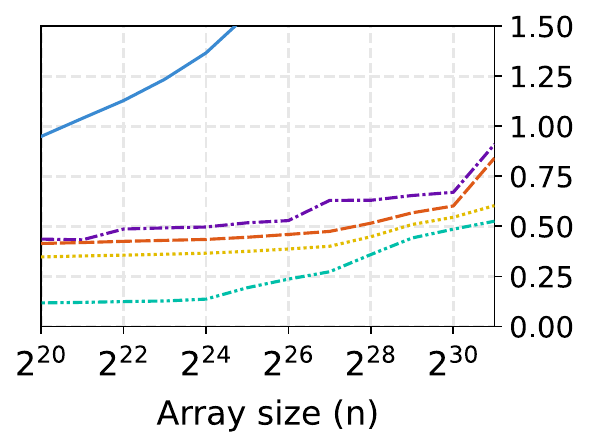}
        \caption{Medium (l, r) range.}
        \label{fig:generations:variations:-2}
    \end{subfigure}%
    \hfill
    \begin{subfigure}[t]{0.345\textwidth}
        \includegraphics[width=\linewidth, clip]{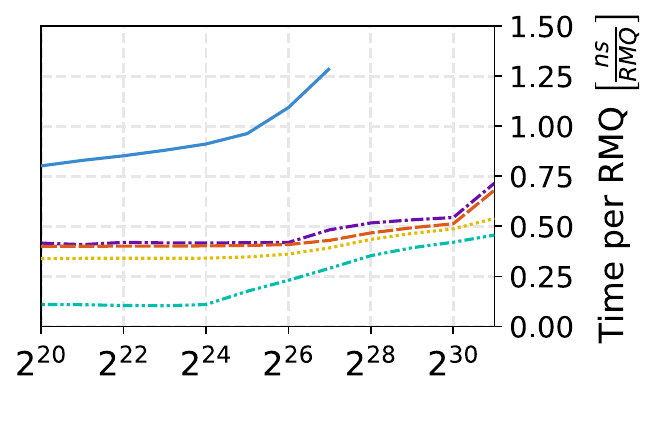}
        \caption{Small (l, r) range.}
        \label{fig:generations:variations:-3}
    \end{subfigure}

    \caption{Step-wise evaluation of the impact of integrating ray tracing into \ours{}.}
    \label{fig:evaluation:plot:time_variations}
\end{figure*}

We observed that smaller values for~$t$ always yield better performance, and therefore, we reduce the number of tests by always selecting~$t$ as small as possible.
For \oursCoalesced{}, this minimal~$t$ is required to be at least twice the chunk size~$c$, whereas for \oursVector{}, $t$~must be greater than or equal to~$c$. 

To identify the best-performing parameter combinations of chunk size~$c$ and group size~$g$, we evaluate a large number of reasonable configurations in \autoref{fig:evaluation:plot:hyperparameter:heatmap} for both  \oursCoalesced{} and \oursVector{} while varying the array size~$n$ from $2^{20}$ to $2^{31}$. For each individual array size, we show the slowdown of each configuration with respect to the best observed configuration of that size.   
We can observe that no single parameter configuration is optimal for all array sizes~$n$. However, for array sizes of up to~$2^{24}$, \oursVector{} consistently achieves the best performance, with an optimal chunk size of $c=2^3$.
Note that \oursVector{} does not allow adjustment of the group size, as it employs vectorized float$4$ loads by default, resulting in an implicit group size of $4$.
For $n > 2^{24}$, \oursCoalesced{} outperforms \oursVector, with slightly varying optimal parameter combinations.
Still, the experiments confirm our prediction from Section~\ref{ssec:core} that choosing the same value for group size and chunk size results in worse performance most of the time.
Overall, a group size of~$g=2^4$ combined with a chunk size of $c=2^5$ yields near-optimal performance across all large array sizes, and therefore, we use this configuration for \oursCoalesced{} in the remainder of the evaluation.
Also note that this aligns the size of a chunk with the size of a GPU cache line, as both are 128~bytes wide.

Unless stated otherwise, we collectively refer to both variants as \ours. For array sizes smaller than $2^{25}$, the reported results correspond to \oursVector, while for larger arrays they correspond to \oursCoalesced.

In addition to these algorithm-specific parameters, we evaluated different CUDA thread block sizes.
Because their impact on execution time was negligible, we use a block size of $1024$ for all GPU-based algorithms throughout the evaluation.

\subsection{Evaluating Optimizations of \ours{}}
\label{ssec:evaluating_optimizations}

\begin{figure}[h!]
    \centering
    \includegraphics[width=\columnwidth, trim={0 0 0 0.9cm}, clip]{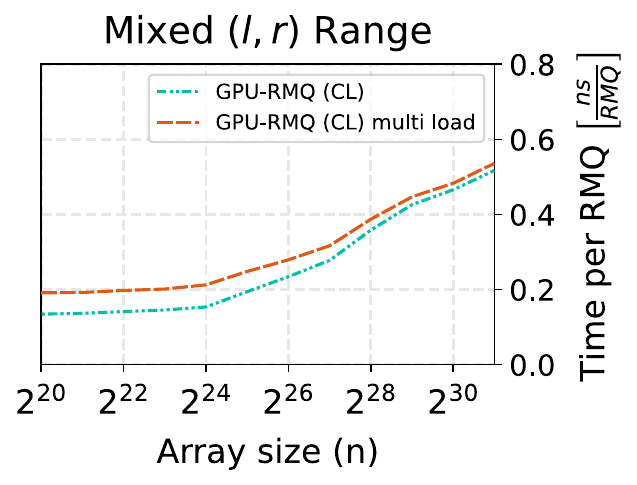}
    \caption{Execution time of \ours{} (CL) for mixed ranges using WLQ compared to the multi-load approach we explained in \autoref{sec:query_assignment}}
    \label{fig:evaluation:plot:multi_load}
\end{figure}

After having tuned the parameters of \ours{}, we experimentally evaluate the best choices for the remaining optimization aspects, namely (a)~whether query assignment should happen via multi-loading or via warp-local queuing (WLQ) and (b)~whether it is beneficial to replace the topmost layer of the hierarchy with the raytracing-based RMQ algorithm.

Regarding (a), Figure~\ref{fig:evaluation:plot:multi_load} compares both query assignment strategies for a batched RMQ task with queries of mixed size.
We vary the size of the input array on the horizontal axis and report the time per RMQ on the vertical axis.
The query batch size, i.e., the number of WLQs to be assigned to threads, remains constant.
As discussed in Section~\ref{sec:query_assignment}, multi-loading performs more memory accesses than WLQ, resulting in higher execution time.
The gap slightly shrinks as the size of the input (and the cost of answering the query) increases, but remains visible.
Consequently, \ours{} defaults to using WLQ.

Unfortunately, WLQ cannot be combined with ray tracing:
NVIDIA requires us to run \ours{} in an OptiX kernel to utilize hardware ray tracing, but because OptiX' programming model can move threads to different processing cores during execution, it might break warp locality guarantees, which is why NVIDIA does not allow warp intrinsics in OptiX.
Consequently, whenever OptiX is involved, we have to rewrite WLQ to a less efficient variant, and switch from warp reductions to atomic aggregation.
We therefore have to investigate whether it is worth losing WLQ and warp aggregation in order to utilize ray tracing, which we integrate into our evaluation of (b). 

The relevant results are shown in Figure~\ref{fig:evaluation:plot:time_variations}:
Alongside \ours{}, we evaluate a version without any warp intrinsics (\textbf{\ours{} w/o warp intrinsics}) and then also run the identical code in an OptiX kernel without using ray tracing (\textbf{\ours{} w/o warp intrinsics in OptiX w/o RT}) to determine the amount of overhead introduced by having to switch to OptiX.
Lastly, we include the hybrid variant where the scan on the topmost layer is replaced by a ray tracing task (\textbf{\ours{} w/o warp intrinsics in OptiX w/ RT}), alongside \RTXRMQ{}, which relies on ray tracing exclusively.

 From the results we can see that executing the code in an OptiX kernel already adds significant overhead, especially for array sizes larger than $2^{30}$.
 Unfortunately, we can also observe that this overhead is not compensated by utilizing hardware ray tracing.
 This shows that using ray tracing for the top layer in the style of~\RTXRMQ{} is currently not recommended, and also explains why \RTXRMQ{} itself is not competitive against our more traditional data structure.
 Ultimately, all following plots do not include the ray tracing variant of \RTXRMQ{}.

\subsection{Memory Footprint}
\label{ssec:memory_footprint}

\begin{figure}[h!]
    \centering
    \includegraphics[width=\columnwidth]{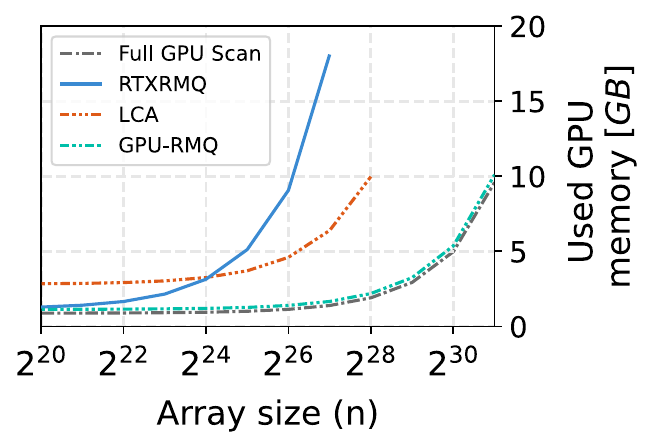}
    \caption{Total memory footprint of all GPU-resident methods.}
    \label{fig:evaluation:plot:total_memory}
\end{figure}

Let us first inspect the memory footprint of \ours{} in comparison to the GPU-resident baseline methods while varying the array size from $2^{20}$ to $2^{31}$. Figure~\ref{fig:evaluation:plot:total_memory} shows the total amount of used GPU memory when executing the respective program, including the original array. 

\begin{figure*}[h!]
    \begin{subfigure}[t]{0.31\textwidth}
        \includegraphics[width=\linewidth, clip]{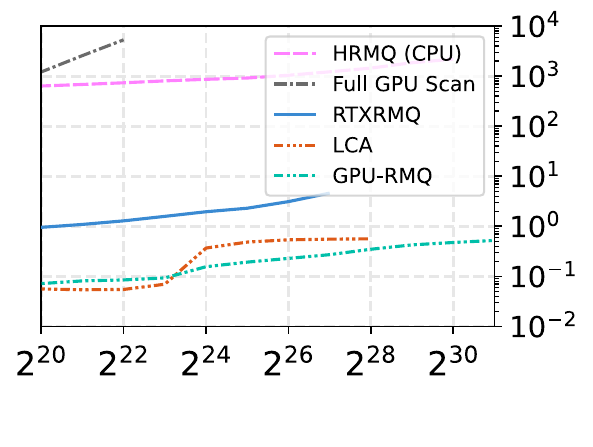}
         \caption{Large (l, r) range.}
        \label{fig:query_time_a}
    \end{subfigure}%
    \begin{subfigure}[t]{0.31\textwidth}
        \includegraphics[width=\linewidth, clip]{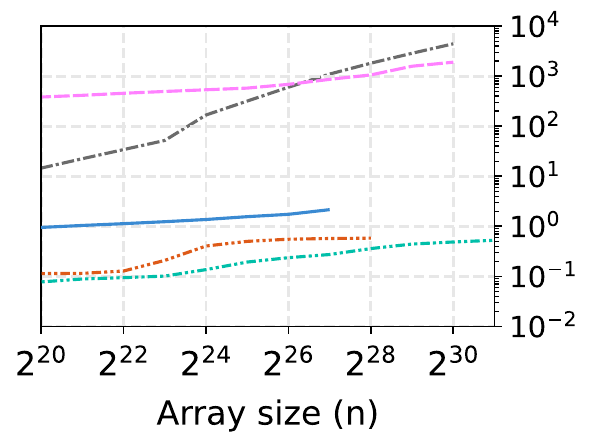}
        \caption{Medium (l, r) range.}
        \label{fig:query_time_b}
    \end{subfigure}%
    \begin{subfigure}[t]{0.345\textwidth}
        \includegraphics[width=\linewidth, clip]{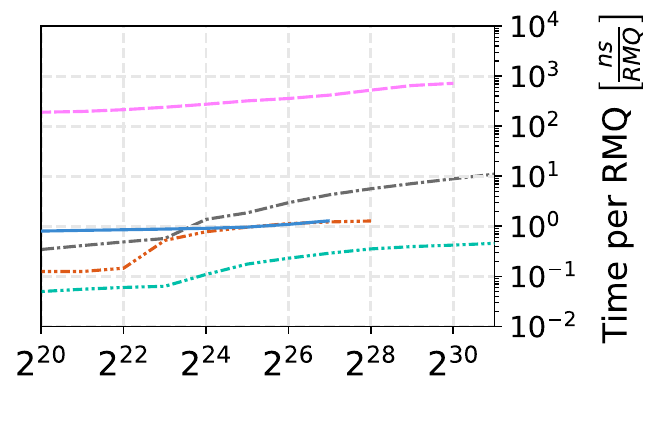}
        \caption{Small (l, r) range.}
        \label{fig:query_time_c}
    \end{subfigure}%

    \caption{Average time per RMQ for a batch of $2^{26}$ queries.}
    \label{fig:evaluation:plot:time_general}
\end{figure*}

Naturally, \exhaustive{} has the minimal memory footprint as it does not build any auxiliary data structures on top of the input array. While \RTXRMQ{} also shows a competitive memory footprint for an array size of $2^{20}$, its memory footprint starts to increase drastically at an array size of $2^{25}$ due the large number of heavy-weight primitives that must be indexed in its BVH. As a result, for \RTXRMQ{}, an array size of $2^{28}$ already exceeds the available GPU memory ($24$GB), rendering it unusable for larger arrays. In comparison, \LCA{} has a higher memory footprint for $2^{20}$ elements than \RTXRMQ{}, which however, increases significantly less with an increase in array size. Nevertheless, array sizes of $2^{29}$ or larger also exceed the available GPU memory.
In contrast, the memory footprint of \ours{} scales gracefully with an increase in array size and positions itself closely above the minimal memory footprint
of \exhaustive{}, since its auxiliary data structure remains small for the configured choices of the chunk size~$c$. Specifically, \ours{} requires at most $30\%$ more GPU memory than \exhaustive{}, whereas \LCA{} consumes up to $5\times$ more and \RTXRMQ{} up to $13\times$ more. As a consequence, \ours{} remains feasible for array size of up to $2^{31}$, being limited by the RTX 4090's main memory size.  
Even on GPUs with larger memory, \ours{} always reaches the same limits as \exhaustive{}, thanks to its very low memory overhead.

\subsection{Construction Time}
\label{ssec:construction_time}

\begin{figure}[h!]
    \centering
    \includegraphics[width=\columnwidth]{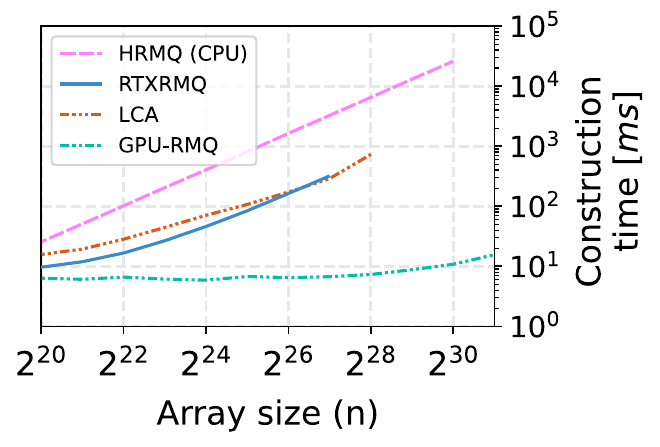}
    \caption{Total construction time for the data structures.}
    \label{fig:evaluation:plot:construction_time}
\end{figure}

Next, we inspect the construction time of the auxiliary data structure of all methods. As \exhaustive{} does not incur any construction time, we do not show it here. Instead, we include the CPU-based variant~\HRMQ{} as a baseline.
Figure~\ref{fig:evaluation:plot:construction_time} shows the construction time in milliseconds on a logarithmic scale for varying input array size. 

From a distance, we can already observe that the construction time of all baseline methods is significantly affected by an increase in array size. In contrast, our \ours{} shows a very stable construction time, worsening only slightly for very large arrays. Moreover, \ours{} archieves the fastest construction time among all methods, outperforming the GPU-resident baselines \RTXRMQ{} and \LCA{}, which perform very similarly. Specifically, \ours{} builds its data structure up to $\sim50\times$ faster than \RTXRMQ{} and up to $\sim100\times$ faster than \LCA. The CPU-based variant \HRMQ{} has the slowest construction time for all input sizes, up to $\sim2400\times$ slower than \ours.

While the construction algorithms of \RTXRMQ{} and \LCA{} scale poorly with an increase in array size, \ours{} naturally utilizes the available parallel resources of the GPU. \RTXRMQ{} suffers from the expensive BVH construction on a large number of primitives, while \LCA{} has to construct a Cartesian tree and perform an Euler tour on it to generate its lookups tables. In contrast, \ours{} constructs each auxiliary layer of its hierarchy from bottom to top in parallel, where a group of $g$~adjacent threads reduces a chunk of $c$~adjacent entries via warp reductions to a single summary. 

\subsection{RMQ Throughput}
\label{ssec:query_time}

After having analyzed the memory footprint and the construction time, let us now inspect the performance for answering RMQs.  Figure~\ref{fig:evaluation:plot:time_general} shows the results, where we report for each method the average time of a single RMQ from the respective batch. 

\begin{figure}[h!]
    \centering
    \includegraphics[width=\columnwidth]{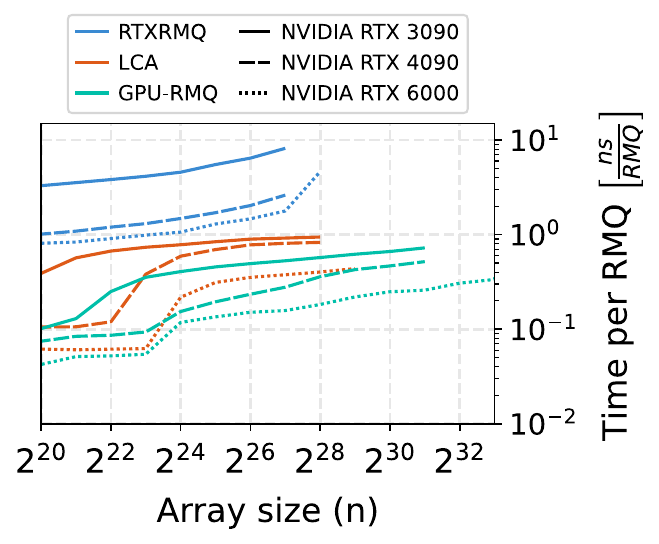}
    \caption{Average time per RMQ for a batch of $2^{26}$ mixed queries, comparing the competitive GPU algorithms across different GPUs.}
    \label{fig:evaluation:plot:architectures}
\end{figure}

First of all, we can see that the CPU-based method \HRMQ{} is several orders-of-magnitude slower than the optimized GPU-based methods for all range sizes. Of all GPU-based methods, \exhaustive{} consistently performs the worst, whereas its performance predictably depends on the range size: The smaller the range, the better it performs. For large and medium ranges and larger array sizes, the runtime per RMQ reaches $10$ms and more, which is not competitive against the remaining methods\footnote{For \exhaustive{}, we stopped the experiment for large and medium ranges and large array sizes as the runtime exceeds any acceptable time frame.}.  
Still, in the best observed case, \exhaustive{} is around one order-of-magnitude slower than the sophisticated GPU-based methods.

\RTXRMQ{} overall performs worse than \LCA{} and \ours{}, especially for large ranges where it performs between $\sim12$ to $\sim17\times$ worse than \ours.
Additionally, it is affected the most by the range size, where handling larger ranges significantly slows it down.
In comparison, both \LCA{} and \ours{} remain relatively unaffected by the range size.
While \LCA{} performs the best for small input arrays, its throughput significantly drops at an array size of $2^{24}$, resulting in up to $\sim8\times$ lower performance than \ours.
This is because \LCA{} performs a small number of random, uncoalesced accesses for processing each query, which results in a memory bottleneck as soon as the working set no longer fits into the GPU's L2 cache.
In contrast, the performance of \ours{} decreases gracefully with an increase in array size, showing the best query performance for the more relevant larger array sizes of $2^{24}$ and larger.




\subsection{Profiling}

To analyze the throughput results in more detail, in the following, we perform a profiling run for the two best methods, namely \LCA{} and \ours{}, in NVIDIA Nsight Compute. Nsight Compute allows for kernel-level profiling of CUDA applications, showing metrics like occupancy, memory transfer rates, instruction throughput, and warp stalls. We run our analysis with $n=2^{28}$ on an RTX 4090 and focus on three reported metrics: (1)~The amount of memory transferred between global memory and L2 cache. (2)~The cache hit rates. (3)~The average number of warp cycles between issued instructions, i.e., the timespan in which the GPU cannot do meaningful work due to memory latency.

In Figure~\ref{fig:evaluation:plot:time_general}, all methods show different behavior for the three RMQ sizes, but \ours{} behaves almost identical across all sizes once $n$ reaches a certain size (around $2^{24}$).
This is surprising at first, given that we would expect small queries to traverse around three layers of the hierarchy, and large ones to traverse five or six, resulting in additional memory loads.
The profiler reveals a 0\% L1 hit rate and 16\% L2 hit rate for small queries, while large queries yield hit rates of 34\% and 61\%, respectively.
Remember that the uppermost layers in the hierarchy have a very low memory footprint, and can easily fit into L2 cache.
As the uppermost layers are cache-resident most of the time while answering large queries, only the lower levels have to be loaded using slow main-memory accesses, resulting in a similar number of main-memory accesses for both large and small RMQs.

Curiously, the profiler also underpins another interesting observation: 
\LCA{} actually performs better when dealing with larger ranges compared to smaller ranges.
For $n=2^{28}$, the \LCA{} loads a total of 21.6~GB from the GPU's main memory when processing the small-range workload, but only 9.8~GB when processing the large-range workload, and the execution time halves accordingly.
As evidenced by Figure~\ref{fig:evaluation:plot:time_general}, this behavior is not an isolated incident, as it repeats across all choices of $n$.
Investigating this effect will be part of future research.

Finally, note that profiling \LCA{} yields between 2500 (large) and 4800 (small) warp cycles per instruction, depending on the range size, meaning that most of the time is spent waiting for random memory accesses, despite the total number of accesses performed by \LCA{} being constant (i.e., independent of the range size).
For comparison, \ours{} exhibits between 18 and 34 average warp cycles per instruction, implying better GPU utilization.

\subsection{Different GPU Generations}
\label{ssec:different_gpu_generations}

So far, we have evaluated all methods on a NVIDIA RTX 4090 GPU from the Ada generation, with 24GB of global memory. Let us now see whether the results look different on an NVIDIA RTX 3090 GPU from the (older) Ampere generation and a NVIDIA RTX 6000 Pro from the (newer) Blackwell generation. Further, an RTX 6000 Pro has 96GB of global memory, and we are interested in seeing how effectively the methods utilize the additional space and which array sizes they can practically handle. 

Figure~\ref{fig:evaluation:plot:architectures} shows the average time per RMQ for \RTXRMQ{}, \LCA{}, and \ours{} on all three GPU generations for the mixed workload. 
First of all, we can see that naturally, the time per RMQ improves for all methods when going from older to newer GPUs. Within each GPU generation, we can see that the relative order between the individual methods remains the same: \RTXRMQ{} is clearly the slowest, \LCA{} performs second best, and \ours{} outperforms the competitors for all array sizes. 
Let us investigate the results for the two newly investigated GPU generation more closely: For the RTX 3090, which only has 6MB of L2 cache, the smallest tested array size of ~$2^{20}$~elements~(4MB) already almost fully occupies the cache. Consequently, for relevant array sizes, all methods suffer from a significant amount of cache misses. However, we can also see that \ours{} can utilize the available cache a bit longer than the competitors, with its performance drop only at $2^{22}$~elements~(16MB).
For the RTX 6000 Pro, which has 96GB of global memory, we can first of all see that all methods can handle larger arrays. However, while \RTXRMQ{} and \LCA{} are only able to double the amount of elements they can index, \ours{} is able to actually handle arrays four times the size. This resembles the increase in available global memory in comparison to the RTX 3090 and RTX 4090. By this, \ours{} is able to handle datasets up to $2^{33}$~elements~(64GB), while \RTXRMQ{} and \LCA{} run out of memory already at $2^{28}$~elements~(1GB) and $2^{29}$~elements~(2GB), respectively.


\section{Conclusion}\label{sec:conclusion}

In this work, we presented \ours{}, a publicly available open-source highly-parallel hierarchical acceleration structure for answering RMQs on GPUs. We showed that it is able to outperform the state-of-the-art CPU and GPU-based alternatives in three essential aspects, namely (a)~memory footprint (up to $\sim4\times$ smaller than \LCA{} and $\sim10\times$ smaller than \RTXRMQ), (b)~construction time (up to $\sim 50\times$ faster than \RTXRMQ{}, $\sim100\times$ faster than \LCA{} and $\sim2400\times$ faster than \HRMQ), and (c)~throughput (up to $\sim 8\times$ faster than \LCA{}, $\sim 17\times$ faster than \RTXRMQ{} and $\sim 4800\times$ faster than \HRMQ) --- enabling RMQ acceleration on GPUs especially for previously not feasible very large datasets. We further demonstrated that \ours{} is highly robust across a variety of different workloads, array sizes, and GPU generations. This new level of acceleration can be beneficial to a wide range of applications in areas including biological sequence analysis, pattern matching, and document retrieval. In addition, our presented techniques can also provide insights for enhancing other scan-heavy general-purpose data structures on GPUs.
We have also investigated the exploitation of heterogeneous resources by combining the usage of CUDA cores and RT cores. Our performance evaluation shows that the associated overheads overshadow potential performance benefits on current generation GPU architectures. However, since RT cores are evolving rapidly, this approach might become more efficient in the near future.


\bibliographystyle{ACM-Reference-Format}
\bibliography{sample}

\end{document}